\documentclass[twocolumn,showpacs,preprintnumbers,amsmath,amssymb]{revtex4}

\usepackage{graphicx}
\usepackage{dcolumn}
\usepackage{bm}
\usepackage{epsfig}
\def\b{\begin{equation}}
\def\e{\end{equation}}

\begin{document}

\title { Electric quadrupole moment of the $4d~^{2}D_{5/2}$ state 
in $\mathrm{^{88}Sr^{+}}$ and its role in an optical frequency standard}

\author {Chiranjib Sur, K. V. P. Latha, Rajat K. Chaudhuri, B. P. Das}
\affiliation{Non-Accelerator Particle Physics Group, Indian Institute of Astrophysics,
Bangalore - 560 034, India}

\author{D. Mukherjee}

\affiliation{Department of Physical Chemistry, Indian Association for the Cultivation of Science, Kolkata -
700 032, India}

\begin{abstract}
The electric quadrupole moment for the $4d~^2D_{5/2}$ state of 
$\mathrm{^{88}Sr^+}$, one
of the most important candidates for an optical clock, has been calculated 
using the relativistic coupled-cluster theory. The result of the calculation 
is presented and the important many-body contributions are
highlighted. The calculated electric quadrupole moment is 
$(2.94\pm0.07)ea_{0}^{2}$, where $a_{0}$ is the Bohr radius and $e$ the 
electronic charge while the measured value is $(2.6\pm0.3)ea_{0}^{2}$.
This is so far the most accurate determination of the electric quadrupole 
moment for the above mentioned state.
\end{abstract}


\maketitle

The frequencies at which atoms emit or absorb electro-magnetic radiation during 
a transition can be used for defining the basic unit of time \cite{time-gen1,time-gen2,
time-gen3}. The transitions that are extremely stable, accurately measurable and
reproducible can serve as excellent frequency standards \cite{time-gen1,time-gen2}. 
The current frequency standard is based on 
the ground state hyperfine transition in $^{133}Cs$ which is in the microwave regime
and has an uncertainty of one part in $10^{15}$ \cite{cs-clock}.
However, demands from several areas of science and technology have lead to a 
worldwide
search \cite{holl} for even more accurate clocks in the optical regime. The 
uncertainty of these clocks is expected to be about 1 part in 
$10^{17}$ or $10^{18}$ \cite{holl}.
Some of the prominent candidates that belong to this category are
$\mathrm{^{199}Hg^+} $ \cite{rafac}, $\mathrm{^{88}Sr^+}$ \cite{bernard,margolis}, 
$\mathrm{^{171}Yb^+}$ 
\cite{stenger} etc. Indeed detailed studies on these ions will have to be carried 
out in order to determine their suitability for optical frequency standards.
In a recent article \cite{phys-world} Gill and 
Margolis have discussed the merits of choosing $\mathrm{^{88}Sr^{+}}$ as a
candidate for an optical clock. Till recently, the most accurate measurement
of an optical frequency was for the clock transition in $\mathrm{^{88}Sr^+}$ which 
has an uncertainty of $3.4$ parts in $10^{15}$ \cite{margolis-sc}.
However, recently, Oskay et al. \cite{hg+_prl} have measured the optical 
frequency 
of $\mathrm{^{199}Hg^+} $ to an accuracy of $1.5$ parts in $10^{15}$ and further 
improvements are expected \cite{It-priv}.

\begin{figure}[h]
  \psfig{figure=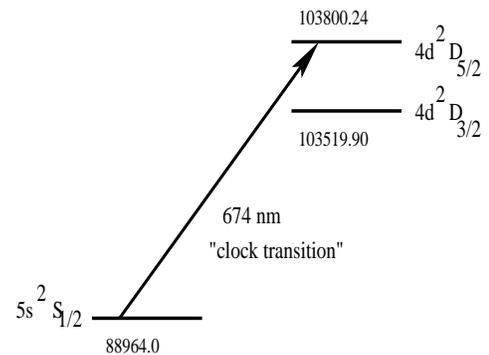,height=4.6cm,width=6.2cm,angle=0}
  \caption{Diagram indicating the clock transition in $\mathrm{^{88}Sr^{+}}$.
           Energy levels are given in $\mathrm{cm^{-1}}$.}
  \label{en-lev}
\end{figure}

When an atom interacts with an external field, the
standard frequency may be shifted from the resonant frequency
\cite{itano-nist}. The quality of
the frequency standard depends upon the accurate and precise measurement
of this shift. To minimize or maintain any shift of the clock frequency, the
interaction of the atom with it's surroundings must be controlled. Hence, it is
important to have a good knowledge of these shifts so as to minimize them while
setting up the frequency standard. Some of these shifts are the linear Zeeman 
shift, quadratic Zeeman shift, second-order Stark shift and electric 
quadrupole shift due to the interaction of atomic electric quadrupole moment 
with the gradient of electric field \cite{itano-nist}. The largest source of 
uncertainty in frequency shift arises from the electric quadrupole 
shift of the clock transition. The electric quadrupole moments of $\mathrm{^{88}Sr^+}$,
$\mathrm{^{199}Hg^+}$ and $\mathrm{^{171}Yb^+}$ have been measured in order to 
determine this shift. However,
accurate measurements of these shifts are very difficult. The calculation of these
shifts although very challenging could in some cases be more accurate than the 
measurements. Such calculations have received relatively less attention so far.
The most rigorous calculation to date has been performed by Oskay et.al 
\cite{hg+_prl} for
$\mathrm{^{199}Hg^+}$ using the relativistic configuration interaction (RCI) method with
a multi-configuration Dirac-Fock - extended optimized level (MCDF-EOL) orbital
basis. We focus on the clock transition 
$5s~^{2}S_{1/2} - 4d~^{2}D_{5/2}$ in $\mathrm{^{88}Sr^+}$ in the present paper. Figure 
\ref{en-lev} shows the clock transition in $\mathrm{^{88}Sr^{+}}$. 
The electric quadrupole moment in the state $4d~^2D_{5/2}$ was measured 
experimentally by Barwood \emph{et al.} at NPL \cite{barwood-srQ}. 
Since the ground state $5s~^{2}S_{1/2}$ does not posses 
any electric quadrupole moment, the contribution to the quadrupole shift for the
clock frequency comes only from the $4d~^{ 2}D_{5/2}$ state.
In this letter, we present our relativistic coupled-cluster (RCC) calculation of the 
electric quadrupole moment of $\mathrm{^{88}Sr^+}$ in the $4d~^2D_{5/2}$ state. RCC is
equivalent to all-order relativistic many-body perturbation theory (RMBPT).

The details of this theory have been discussed in several
articles \cite{bishop,bartlett}. Here we shall only give a brief outline. Treating the
closed shell Dirac-Fock (DF) state $\left|\Phi\right\rangle $ as the reference state,
the exact wave function in RCC theory can be expressed as,
\begin{equation}
\left|\Psi\right\rangle =\exp(T)\left|\Phi\right\rangle ,
\label{cc-1}
\end{equation}
where $T$ is the core electron excitation operator. 
In the coupled cluster singles and doubles (CCSD)
approximation, $T$ can be expressed as the sum
of one- and two-body excitation operators, \emph{i.e.} $T=T_{1}+T_{2}$,
and can be written in the second quantized form as
\begin{equation}
T=T_{1}+T_{2}=\sum_{ap}a_{p}^{\dagger}a_{a}t_{a}^{p}+\frac{1}{2}\sum_{abpq}a_{p}^{\dagger}a_{q}^{\dagger}a_{b}a_{a}t_{ab}^{pq}.
\label{cc-7}
\end{equation}
where $t_a^p$ and $t_{ab}^{pq}$ are the amplitudes of the singles and double
excitation operators respectively. The normal ordered Hamiltonian can be written as,
\begin{equation}
H_{N}\equiv H-\left\langle \Phi\right|H\left|\Phi\right\rangle =H-E_{DF},
\label{cc-4}
\end{equation}
where $H$ is the Dirac-Coulomb Hamiltonian given by
\begin{equation}
  H_{DC} = \sum_{i}^{N} \left [c \mbox{\boldmath{$\alpha_i$}} \cdot. \mbox{\boldmath $p_i$}
+{\bf\beta_i} c^2+ V(r_i)\right ] + \sum_{i<j} \frac{1}{r_{ij}}
\end{equation}
For  a single valence system we define the reference state as,
\begin{equation}
\left|\Phi_{v}^{N+1}\right\rangle \equiv a_{v}^{\dagger}\left|\Phi\right\rangle 
\label{cc-10}
\end{equation}
with the particle creation operator $a_{v}^{\dagger}$. The wave function corresponding
to this state can be written as,
\begin{equation}
\left|\Psi_{v}^{N+1}\right\rangle =\exp(T)\left\{ \left(1+S_v\right)\right\} \left|\Phi_{v}^{N+1}\right\rangle .
\label{cc-11}
\end{equation}
where,
\begin{equation}
S_{v}=S_{1v}+S_{2v}=\sum_{v\neq p}a_{p}^{\dagger}a_{v}s_{v}^{p}+\sum_{bpq}a_{p}^{\dagger}a_{q}^{\dagger}a_{b}a_{v}s_{vb}^{pq}\,,
\label{s-expr}
\end{equation}
where $S$ corresponds to the excitation operator in the valence sector and 
$v$ stands for valence orbital and $s_v^p$ and $s_{vb}^{pq}$ are the amplitudes
of single and doubles excitations respectively.
Details concerning the evaluation of the closed and open shell amplitudes have been
discussed earlier \cite{geetha}.
Triple excitations are included in our open shell RCC amplitude calculations
in an approximate way(CCSD(T))\cite{ccsd(t)}.

The expectation value of any operator $O$ with respect to the state 
$\left|\Psi^{N+1}\right\rangle $ is given by,

\begin{eqnarray}
&&\left\langle O\right\rangle =\frac{\left\langle \Psi^{N+1}\right|O\left|\Psi^{N+1}\right\rangle }{\left\langle \Psi^{N+1}\right|\left.\Psi^{N+1}\right\rangle }\nonumber\\
&&=\frac{\left\langle \Phi^{N+1}\right|\left\{ 1+S^{\dagger}\right\} \bar O \left\{ 1+S\right\} \left|\Phi^{N+1}\right\rangle }{\left\langle \Phi^{N+1}\right|\left\{ 1+S^{\dagger}\right\} \exp(T^{\dagger})\exp(T)\left\{ 1+S\right\} \left|\Phi^{N+1}\right\rangle }.
\label{cc-16}
\end{eqnarray}
where $\bar O = \exp(T^\dagger) O \exp(T)$.

\begin{figure}
  \psfig{figure=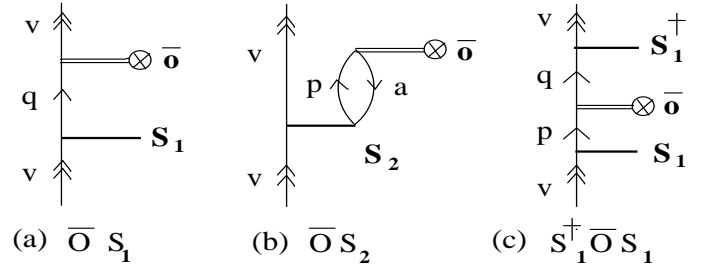,height=3.5cm,width=9.0cm,angle=0}
  \caption{The diagrams (a) and (c) are subsets of dressed pair correlation 
           (DPC) diagrams. Diagram (b) is one of the direct dressed 
           core-polarization (DCP) diagram.}
  \label{obard}
\end{figure}

The first few terms in the above expectation value
can be identified as $\overline{O}$, $\overline{O}S_{1}$, 
$\overline{O}S_{2}$, $S_{1}^\dagger\overline{O}S_{1}$ etc; 
are referred to as dressed Dirac-Fock (DDF), dressed pair correlation (DPC) 
(Fig.\ref{obard}(a)) and dressed core polarization (DCP)(Fig.\ref{obard}(b)) 
respectively. We use the term `dressed' 
because the operator $O$ includes the effects of the core excitation 
operator $T$. Among the above, we can identify few other terms which 
play crucial role in determining the correlation effects. One of those terms 
is $S_{1}^\dagger\overline{O}S_{1}+c.c$ ((Fig.\ref{obard}(c)) 
which is called as dressed higher order pair
correlation (DHOPC) since it directly involves the correlation between a
pair of electrons. In table \ref{contrbn} individual contributions from these
diagrams are listed. 

The orbitals used in the present work are expanded in terms of a finite
basis set comprising of Gaussian type orbitals (GTO) \cite{napp-fbse}
\begin{equation}
F_{i,k}(r)=r^{k}\exp(-\alpha_{i}r^{2}),
\label{comp-1}
\end{equation}
with $k=0,1,2\cdots$ for $s,p,d,\cdots$ type functions, respectively.
The exponents are determined by the even tempering condition \cite{even-tem}
\begin{equation}
\alpha_{i}=\alpha_{0}\beta^{i-1}.
\label{comp-2}
\end{equation}
The starting point of the computation is the generation of the Dirac-Fock
(DF) orbitals \cite{napp-fbse} which are defined on a radial grid
of the form
\begin{equation}
r_{i}=r_{0}\left[\exp(i-1)h-1\right]
\label{comp-3}
\end{equation}
with the freedom of choosing the parameters $r_{0}$ and $h$. All
DF orbitals are generated using a two parameter Fermi nuclear distribution
\begin{equation}
\rho=\frac{\rho_{0}}{1+\exp((r-c)/a)},
\label{fermi-nucl}
\end{equation}
 where the parameter $c$ is the half charge radius and $a$ is related
to skin thickness, defined as the interval of the nuclear thickness
in which the nuclear charge density falls from near one to near zero.
The interaction of the atomic quadrupole moment with  the external electric-field gradient 
is analogous to the interaction of a nuclear quadrupole moment with the 
electric fields generated by the atomic electrons inside the nucleus. 
In the presence of the electric field, this gives rise to an energy 
shift by coupling with the gradient of the electric field.
Thus the treatment of electric quadrupole moment is analogous to 
the nuclear counterpart. The quadrupole moment ${\bf \Theta}$ of an atomic
state $|\Psi(\gamma,J,M)\rangle$ is defined as the diagonal matrix element of the
quadrupole operator with the maximum value $M_J$, given by,
\begin{equation}
  {\bf \Theta}= \langle \Psi(\gamma ~J~J)|\Theta_{zz}|\Psi(\gamma ~J~J)\rangle.
\label{Qij}
\end{equation}
Here $\gamma $ is an additional quantum number which distinguishes the initial 
and final states. The electric quadrupole operator in terms of the electronic
coordinates is given by,
$$ \Theta_{zz} = -\frac{e}{2} \sum_j \left(3z_j^2 - r_j^2\right) ,$$
where the sum is over all the electrons and  $z$ is the coordinate of the  
$j$th electron.
To calculate the quantity we express the quadrupole operator in its single 
particle form as
\begin{equation}
\Theta_{m}^{(2)}=\sum_{m}q_{m}^{(2)}
\end{equation}
and the single particle reduced matrix element is expressed as \cite{grant-rad}
\begin{widetext}
\begin{equation}
\left\langle j_{f}\right\Vert q_{m}^{(2)}\left\Vert j_{i}\right\rangle =\left\langle j_{f}\right\Vert C_{m}^{(2)}\left\Vert j_{i}\right\rangle \frac{15}{k^{2}}
\int dr\left\{ j_{2}(kr)\left(P_{\kappa_{f}}P_{\kappa_{i}}
+Q_{\kappa_{f}}Q_{\kappa_{i}}\right)+j_{3}(kr)\frac{\left(j_{j}-j_{i}-1\right)}{3}\left(P_{\kappa_{f}}Q_{\kappa_{i}}+Q_{\kappa_{f}}P_{\kappa_{i}}\right)\right\}. 
\label{quad-eq}
\end{equation}
\end{widetext}
In Eq.($\ref{quad-eq}$), the subscripts $f$ and $i$ correspond to the final and 
initial states respectively and $j_{m}(kr)$ is the Bessel 
function of order $m$; $P$ and $Q$ are the radial part of the large and 
small components of the single particle Dirac-Fock wavefunctions respectively.
$j_{i}$ is the total angular momentum and $\kappa_{i}$ is the relativistic 
angular momentum quantum number for the $i$th electron.
The angular factor is given by
\begin{eqnarray}
\left\langle j_{f}\right\Vert C_{m}^{(k)}\left\Vert j_{i}\right\rangle =
&&(-1)^{(j_{f}+1/2)}\sqrt{(2j_{f}+1)}\sqrt{(2j_{i}+1)}\nonumber\\
&&\times\left(\begin{array}{ccc}
j_{f} & 2 & j_{i}\\
-1/2 & 0 & 1/2\end{array}\right)\pi(l,k,l^{\prime})
\label{ang}
\end{eqnarray}
where 
\[
\pi(l,k,l^{\prime})=\left\{ \begin{array}{c}
\begin{array}{cc}
1 & \mathrm{if}\: l+k+l^{\prime}\,\,\mathrm{even}\\
0 & \mathrm{otherwise}\end{array}\end{array}\right.\]
$l$ and $k$ being the orbital angular momentum and the rank respectively.

Finally using the Wigner Eckart theorem we define the electric quadrupole 
moment in terms of the reduced matrix elements as 
\begin{equation}
\left\langle j_{f}\right| \Theta_{m}^{(2)} \left| j_{i}\right\rangle=(-1)^{j_{f}-m_{f}}\left(\begin{array}{ccc}
j_{f} & 2 & j_{i}\\
-m_{f} & 0 & m_{f}\end{array}\right)
\left\langle j_{f}\right\Vert \Theta^{(2)}\left\Vert j_{i}\right\rangle
\label{wig-eck}
\end{equation}

In table \ref{gauss-table} we have presented the details of the basis
functions used in this calculation and in table  \ref{contrbn} contributions
from different many body terms.
The value of ${\bf \Theta}$ in the $4d~^{2}D_{5/2}$ state measured experimentally is
$(2.6\pm 0.3)ea_0^{2}$ \cite{barwood-srQ}, where $e$ is the electronic charge and 
$a_0$ is the Bohr radius. Our calculated value for the 
$4d~^{2}D_{5/2}$ stretched state is $(2.94\pm0.07)ea_{0}^{2}$.
We have estimated the error incurred in our present work, by taking the 
difference between our RCC calculations with singles, doubles as well as the most
important triple excitations (CCSD(T)) and only single and double 
excitations (CCSD).
A non-relativistic Hartree-Fock (HF) determination resulted in 
${\bf \Theta}=3.03ea_{0}^{2}$ \cite{barwood-srQ}. A subsequent calculation based 
on RCI with MCDF-EOL orbital basis yielded ${\bf \Theta}=3.02 ea_{0}^{2}$ 
\cite{hg+_prl}. In that calculation, correlation effects 
arising from a subset of the terms $S_1$, $T_1$ for single excitations and 
$S_2$ and $T_2$ for double excitations were considered, where $S_1$ and $S_2$ are
the cluster operators representing single excitations from the valence
$5s$ orbital to a virtual orbital and double excitations from the valence $5s$ and the core
$\{4s,~ 4p,~ 3d\}$ orbitals, with atmost one excitation from the core, respectively.
In our calculation, in addition to these effects, the effects arising from the 
non-linear terms like $T_2^2$, $T_1T_2$, etc. have been included.
In the framework of CCSD theory, the single
and double excitations have been treated to all orders in electron correlation
including excitations from the entire core. This amounts to a more rigorous 
treatment of electron correlation in comparison to the previous calculation performed
using the RCI method..

It is clear from the table \ref{contrbn} that the DDF contribution
is the largest. The leading correlation contribution comes from the DPC effects
and the DCP effects are an order of magnitude smaller. This can be understood
from the DPC diagram (Fig.\ref{obard}(a)) which has a valence electron 
in the $4d_{5/2}$ state. Hence the dominant contribution to the electric quadrupole 
moment arises from the overlap between virtual $d_{5/2}$ orbitals and the valence, 
owing to the fact that $S_1$ is an operator of rank 0 and 
the electric quadrupole matrix elements for the valence $4d_{5/2}$ and the diffuse
virtual $d_{5/2}$ orbitals are substantial.
On the other hand, in the DCP diagram (Fig.\ref{obard}(b)), the matrix 
element of the same operator could also involve the less diffuse $s$ or $p$ 
orbitals. Hence, for a property like the electric quadrupole moment, whose 
magnitude depends on the square of the radial distance from the nucleus, 
this trend seems reasonable, whereas for properties like hyperfine interaction
which is sensitive to the near nuclear region, the trend is just the opposite 
for the $d$ states \cite{csur-mg+}. As expected, the contribution of the DHOPC 
effect 
i.e., $S_1^\dagger \bar O S_1 $ (Fig.\ref{obard}(c)) is relatively important as it 
involves an electric quadrupole matrix element between the valence $4d_{5/2}$ and a
virtual $d_{5/2}$ orbital. Unlike many other properties particularly the hyperfine 
interactions, an all-order determination of this diagram is essential for obtaining an
accurate value of the electric quadrupole moment. One of the strengths of RCC theory is 
that it can evaluate such diagrams to all orders in the residual Coulomb interaction.
\begin{table}
\caption{\label{gauss-table}No. of basis functions used to generate the even
tempered Dirac-Fock orbitals and the corresponding value of $\alpha_{0}$
and $\beta$ used.}
~
\begin{center}\begin{tabular}{llllllllll}
\hline 
&
$s_{1/2}$&
$p_{1/2}$&
$p_{3/2}$&
$d_{3/2}$&
$d_{5/2}$&
$f_{5/2}$&
$f_{7/2}$&
$g_{7/2}$&
$g_{9/2}$\tabularnewline
\hline
\hline 
Number of basis&
38&  35& 35& 30& 30& 25& 25& 20&  20
\tabularnewline
$\alpha_{0}(\times10^{-5})$&
525& 525& 525& 425& 425& 427& 427& 425& 425
\tabularnewline
$\beta$&
2.33& 2.33& 2.33& 2.13& 2.13& 2.13& 2.13& 1.98& 1.98
\tabularnewline
Active holes&
10& 10& 10& 11& 11& 8& 8& 6& 6
\tabularnewline
Active particles&
4& 3& 3& 1& 1& 0& 0& 0& 0
\tabularnewline
\hline
\hline 
& & & & & & & & & 
\tabularnewline
\end{tabular}
\end{center}
\end{table}
\begin{table}
\caption{\label{contrbn}Contributions of the electric quadrupole moment in
atomic units from different many-body effects in
the CCSD(T) calculation. The terms like DDF, DCP, DPC, DHOPC are
explained in the text. The remaining terms in Eq.$\ref{cc-16}$ are referred 
to as `others'.}
\begin{center}
\begin{tabular}{llllll}
\hline 
   DDF&         DPC&         DCP&    DHOPC&      Others&   Total
\tabularnewline
\hline
\hline
\tabularnewline
   3.4963&   -0.4306&    -0.0642&   0.0353&        -0.0271&  2.94   \\
          &         &          &          &              &           \\
\hline 
\end{tabular}
\end{center}
\end{table}

In conclusion, we have performed an \emph{ab initio} calculation of the 
electric quadrupole moment for the  $4d~^2 D_{5/2}$ state of 
$\mathrm{^{88}Sr^+}$ to an accuracy of 
less than 2.5 \%  using the RCC theory. Evaluation of correlation effects to 
all orders as well as the inclusion of the dominant triple excitations in 
our calculation was crucial in achieving this accuracy. This is currently
the most accurate determination of electric quadrupole moment for this state in
$Sr^{+}$. Our result will lead 
to a better quantitative understanding of the electric quadrupole shift of 
the resonance frequency of the clock transition in $\mathrm{^{88}Sr^+}$.

This work was supported by the BRNS for project no. 2002/37/12/BRNS. The 
computations were carried out on our group's Xeon PC cluster.
We are grateful to Dr. Wayne Itano and Dr. Geoffrey Barwood, 
for helpful discussions.


\end{document}